\documentclass[prb,aps,twocolumn]{revtex4-1}

\usepackage{amsmath}  
\usepackage{amsfonts} 
\usepackage{graphicx} 
\usepackage{color}
\usepackage{import}
\usepackage{listings}
\usepackage{color}
\usepackage{url}

\definecolor{dkgreen}{rgb}{0,0.6,0}
\definecolor{gray}{rgb}{0.5,0.5,0.5}
\definecolor{mauve}{rgb}{0.58,0,0.82}

\begin{document}

\title{Undergraduate Quantum Mechanics: a Numerical Approach using QuTiP}

\author{Andrew M.C. Dawes}
\email{dawes@pacificu.edu} 
\affiliation{Department of Physics, Pacific University, Forest Grove, OR 97116}

\date{\today}

\begin{abstract}
  We present an introduction to the Quantum Toolbox in Python (QuTiP) in the context of an undergraduate quantum mechanics class and potential senior research projects. QuTiP provides ready-to-use definitions of standard quantum states and operators as well as numerous dynamic solvers and tools for visualization. The quantum systems described here are typical for an undergraduate curriculum and range from two-state systems to optical interaction with multilevel atoms.
\end{abstract}

\maketitle 

\section{Introduction} 
\label{sec:intro}

Quantum mechanics has undergone substantial pedagogical evolution in response to Physics Education Research (PER) results and advances in laboratory technology.\cite{Beck:2012aa} The development of a specialized framework for computational calculation in quantum mechanics provides a valuable addition that we encourage instructors to add to their courses. Our goal is to present several ways computational techniques can be added to existing courses in quantum mechanics. A distinct advantage of this approach is that the computational framework includes research-grade tools that enable advanced problem solving for senior undergraduates and even graduate students.

We note that this approach may be most natural for courses that emphasize matrix mechanics early in the course as vector and matrix methods are the most natural representations in numerical work. The order of the specific examples presented here follows the textbook \emph{Quantum Mechanics: Theory and Experiment} by Mark Beck \cite{Beck:2012aa} which starts with photon polarization states and spin states then moves to one-dimensional potentials and time dependent states. Many courses follow a similar progression, and the tools we describe are useful in other contexts as well. Even a course starting with wave mechanics can make use of these examples once spin states and matrix methods are introduced.

We present results using a specific computational framework called QuTiP (the Quantum Toolbox in Python). The QuTiP computational framework was released in 2011 with additional versions and updates in the years following.\cite{Johansson:2012aa,Johansson:2013aa} The examples here have been developed using version 4.2. In addition to QuTiP, we recommend using Jupyter notebooks which provide an interactive python computing environment.\cite{Kluyver:2016aa,Jupyter} The notebook interface allows instructors to create interactive lessons and for students to document their work in-place.

\section{Photon polarization states}
\label{sec:photons}
Our first example system uses photon polarization states in parallel with the text and associated experiments presented by Beck.\cite{Beck:2012aa} Given the similarity of the matrix representation of all two-state systems, users benefit from code reuse and conceptual recycling when moving between these topics.

Photon polarization states are represented by vectors in an 2-dimensional space. Starting with small vectors and matrices allows simultaneous calculation by-hand and via computer. In the context of teaching, this is useful for verification during the orientation phase and to catch numerical errors while students learn a new system. Later, students develop their own habits of verification and learn ways to check their work for larger Hilbert spaces.

Python is an object-oriented language which for our purposes means that quantum states are represented by an object. Other pre-defeined functions can operate on these objects, and we can define our own operations as discussed below. The most fundamental object in QuTiP is the \verb|Qobj|. Instances of \verb|Qobj| have many useful properties. Primarily, they are a matrix (or vector) implemented efficiently in Python. The features most relevant to this work are the indication of their status as either bra or ket, and whether they are Hermitian operators or not. We will not make much use of the other properties, and for the most part, these properties are used internally for type-checking and other verification. Several important functions create frequently-used quantum objects. One of the most fundamental functions creates basis vectors. The function \verb|basis| takes a dimension, $N$, and an index as its arguments. For two-state systems, $N=2$ and we have two basis vectors: \verb|basis(2,0)| and \verb|basis(2,1)| which represent the column vectors $\begin{pmatrix}
  1 \\ 0
\end{pmatrix}$ and $\begin{pmatrix}
  0 \\ 1
\end{pmatrix}$ respectively.\footnote{Python is zero-indexed so the first element in an array or list has an index value of 0.} We demonstrate these and many other features in the following sections.

Before presenting sample code, it is important to note that while a stock python environment is quite functional, it does not include some features we use throughout the following examples. To add these features, we import from the \verb|qutip|, \verb|numpy|, and \verb|matplotlib| packages:\cite{Oliphant:2015aa,Jones:2001aa,Hunter:2007aa}

\begin{lstlisting}[caption={Module imports used for code samples presented.},label={lst:imports}]
from qutip import *
from numpy import sqrt, pi, array, sin, cos, arange
import matplotlib.pyplot as plt
%matplotlib inline
\end{lstlisting}
The last line is only useful in a Jupyter Notebooks, it enables inline figures.

For a two-level system, the basis states themselves can serve as horizontal ($|H\rangle$) and vertical ($|V\rangle$) polarization states or as spin-up ($|+z\rangle$) and spin-down ($|-z\rangle$). For now, we use polarization states. In parallel with the notation of Beck \cite{Beck:2012aa}, we can define three pairs of polarization states via code shown in Listing~\ref{lst:photons}. Note, these are all written in the HV basis; a point we make as it is important to be aware of the basis any time a vector or matrix is used to represent a quantum state.\footnote{Work is underway to add basis information to Qobj.}

\begin{lstlisting}[caption={Definitions of basis states for the three standard photon polarizations represented in the HV-basis.},label={lst:photons}]
H = basis(2,0)
V = basis(2,1)
P45 = 1/sqrt(2)*(H + V)
M45 = 1/sqrt(2)*(H - V)
R = 1/sqrt(2)*(H - 1j*V)
L = 1/sqrt(2)*(H + 1j*V)
\end{lstlisting}

Photon polarization states are a particularly good starting point in the pedagogical sense as students can relate their understanding to observations from optics experiments and the vector nature of polarization. With these states defined, we can explore basic matrix mechanics. Inner products such as $\langle H | V \rangle$ are computed by converting the ket $|H\rangle$ into the bra $\langle H |$ using the dagger \verb|.dag()| method which is defined for all instances of \verb|Qobj|. As we would expect for orthogonal vectors, the inner product $\langle H | V \rangle = 0$. In QuTiP/python, \verb|H.dag()*V| returns 0.

We make futher use of the \verb|.dag()| method to create projection operators:
\begin{align}
  \hat{P}_H &= |H\rangle\langle H| \\
  \hat{P}_V &= |V\rangle\langle V|,
  \label{eqn:proj}
\end{align}
which can be used to project a state into either horiziontal or vertical polarization. In this way, they represent physical operation of passing light through a polarizer. The QuTiP implementation of these projection operators is given in Listing~\ref{lst:proj}. As a simple calculation example, one can calculate $\hat{P}_V |\psi_{+45}\rangle$ which represents the photon state that would result from passing light polarized at $+45^\circ$ through a vertical polarizer. The computational representation is as simple as declaring \verb|psi = P45| and operating on \verb|psi| with \verb|Pv|: \verb|Pv*psi| yields $\begin{pmatrix}
  0 \\ 0.707
\end{pmatrix}$,
as expected.

\begin{lstlisting}[caption={Projection operators for horizontal and vertical polarization.},label={lst:proj}]
Ph = H*H.dag()
Pv = V*V.dag()
\end{lstlisting}

Another example of a simple matrix operation that is readily implemented in QuTiP is the polarization rotation matrix. This operator corresponds to the rotation of a polarization state by a given angle. Defined as
\begin{equation}
\hat{R}_p(\theta) = \begin{pmatrix} \cos\theta & -\sin\theta \\ \sin\theta & \cos\theta \end{pmatrix},
\end{equation}
the rotation matrix implementation is given in Listing~\ref{lst:rot}. We use a short python function to create the matrix for a given angle \verb|theta|. The function returns a \verb|Qobj| that has been processed with the \verb|.tidyup()| which removes any very small elements from the returned \verb|Qobj|. This is often useful as numerical artifacts from the finite precision of the computer may accumulate if operators are used repeatedly.\footnote{Such artifacts can be seen by calculating \texttt{cos(pi/2)} which evaluates to 1e-17 rather than 0. The \texttt{.tidyup()} method rounds such values back to zero.}

\begin{lstlisting}[caption={2x2 polarization rotation operator.},label={lst:rot}]
def Rp(theta):
    M = [[cos(theta),-sin(theta)],
         [sin(theta),cos(theta)]]
    return Qobj(M).tidyup()
\end{lstlisting}

The photon states are also a good place to practice matrix operations such as change-of-basis. QuTiP handles matrix operations in a straightforward way making it easy to perform transformation operations. As an example, we define the similarity transform used to change from one basis to another (Listing~\ref{lst:simtrans}). This is another example of a python function written to generate a matrix. In this case, we generate the matrix that transforms a state from one basis to another. As inputs, the function takes four vectors: the two old basis vectors and the two new basis vectors. Looking inside the function, we see an example of an outer product: \verb|new1.dag() * old1|. The \verb|.dag()| again method converts a ket to a bra in order to carry out inner products such as: $\langle new_1 | old_1 \rangle$. The last two lines of the function extract the value of each of the four inner products \verb|a,b,c,d| and assemble them into a \verb|Qobj| matrix. The full output of this function is the matrix
\begin{equation}
  \bar{S}\doteq \begin{pmatrix}
    \langle new_1 | old_1 \rangle & \langle new_1 | old_2 \rangle\\
    \langle new_2 | old_1 \rangle & \langle new_2 | old_2 \rangle
  \end{pmatrix}.
\end{equation}

\begin{lstlisting}[caption={Definitions of basis states for the three standard photon polarizations represented in the HV-basis.},label={lst:simtrans}]
def sim_transform(old1, old2, new1, new2):
    '''
    Form the similarity transform from one
    basis to another.
    '''

    # Calculate the relevant inner products:
    a = new1.dag()*old1
    b = new1.dag()*old2
    c = new2.dag()*old1
    d = new2.dag()*old2

    # extract values from these Qobj inner
    # product results, reshape to 2 by 2, and
    # form a new Qobj matrix:
    s = [i.data[0,0] for i in [a,b,c,d]]
    return Qobj(array(s).reshape(2,2))
\end{lstlisting}

As we show, QuTiP can be used to construct many operators that are typically developed and used by hand. While the 2x2 cases are straightforward, we find that once users are familiar with the \verb|Qobj| and the various methods provided by QuTiP, they can readily extend the size of the Hilbert space to scale their work to more complex problems. In the following sections, we describe several such examples; additional intermediate and advanced examples are available via the QuTiP website and documentation.\footnote{\protect\url{www.qutip.org}}

\section{Spin-1/2}

At the level shown thus far, QuTiP can serve as a matrix-mechanics calculator, used to compute vector and matrix calculations in a way that follows Dirac notation and respects the algebra of quantum states. While such a use is welcome in the teaching setting, QuTiP provides users with much more. Here, we demonstrate one of the QuTiP equation solvers. We also transition to spin-1/2 system and define the six relevant states as in Listing~\ref{lst:spins}. Note, the same QuTiP function \verb|basis(...)| is used again here because spin-1/2 states are also represented by 2-element column vectors just as polarization states were in Section~\ref{sec:photons}.

\begin{lstlisting}[caption={The spin-1/2 system states in the z-basis using a variable naming convention where ``p'' indicates ``+'' and ``m'' indicates ``-'' and the three coordinate directions are given by the second letter in each variable name (i.e. \texttt{mx} is $|-x\rangle$ ). We note that these are all kets as defined here.},label={lst:spins}]
pz = basis(2,0) # i.e. the column vector (1,0)
mz = basis(2,1) # i.e. the column vector (0,1)
px = 1/sqrt(2)*(pz + mz)
mx = 1/sqrt(2)*(pz - mz)
py = 1/sqrt(2)*(pz + 1j*mz)
my = 1/sqrt(2)*(pz - 1j*mz)
\end{lstlisting}

The spin-1/2 states are written in the basis where \verb|basis(2,0)| creates the vector we use to represent spin-up along the $z$-axis: $$|+z\rangle \doteq\begin{pmatrix} 1 \\ 0 \end{pmatrix}.$$ The declaration of this and the other spin states is given in Listing~\ref{lst:spins}.

In contrast to photon polarization states, where the orthogonality of $|H\rangle$ and $|V\rangle$ is not surprising, it is important to verify the orthogonality of the spin-1/2 states. The idea that the $|+z\rangle$ state is orthogonal to the $|-z\rangle$ state may be quite counterintuitive given the names spin-up and spin-down: ``up'' and ``down'' are classically anti-parallel rather than orthogonal. Fortunately, vector calculation by hand and via \verb|mz.dag()*pz| confirm the $z$ states are orthogonal. This is easily generalized for $x$ and $y$. To drive the point home, we find that $|+z\rangle$ and $|+y\rangle$ are \textbf{not} orthogonal (as one may expect when thinking strictly of the cartesian unit vectors). Nonetheless, \verb|pz.dag()*py| confirms the correct result is 0.707 ($\frac{1}{\sqrt{2}}$).\footnote{At this point, we call out another benefit of including numerical work in a quantum mechanics class, and that is encouraging students to identify decimal representations of numerical factors common to vector projection such as 0.707}.

\begin{lstlisting}[caption={The spin-1/2 projection operators are readily defined in terms of the Pauli matrices.},label={lst:spinops}]
Sx = 1/2.0*sigmax()
Sy = 1/2.0*sigmay()
Sz = 1/2.0*sigmaz()
\end{lstlisting}

Similar to the spin-1/2 basis states, the Pauli matrices are available in QuTiP: \verb|sigmax|, \verb|sigmay|, and \verb|sigmaz|. The three spin operators are given in Listing~\ref{lst:spinops} where we take $\hbar = 1$, a step that should made quite clear in the classroom! With these objects defined, it is straightforward to construct a Hamiltonian that corresponds to an external magnetic field; a system with rich dynamics that can still be computed by hand for comparison. Listing~\ref{lst:bfieldH} demonstrates the creation of a Hamiltonian that corresponds to a magnetic field oriented in the $+z$ direction: $$H= - \mathbf{\mu}\cdot \mathbf{B} =-\gamma S_z B,$$ with the $B$-field taken as a classical field, the Hamiltonian operator is:
\begin{equation}
  \hat{H_z} = -\Omega \hat{S}_z.
  \label{eqn:Hz}
\end{equation}

\begin{lstlisting}[caption={Code to represent the Hamiltonian for a single spin in a magnetic field oriented in the $+z$ direction.},label={lst:bfieldH}]
omega = 5
Hz = -omega*Sz
\end{lstlisting}

First, we point out the visual similarity between Eq.~\ref{eqn:Hz} and the python code to construct the same Hamiltonian \verb|Hz=-omega*Sz|. This similarity demonstrates QuTiP's ease-of-use and illustrates the readable nature of the relevant code. Next, the evolution of spin-1/2 particles subject to Eqn.~\ref{eqn:Hz} is calculated by setting an initial state for the system, and allowing the system to evolve according to the time-dependent Schr{\"o}dinger equation. The solver used in this case is \verb|sesolve()| which takes at least four arguments: the Hamiltonian (\verb|Hz|), the initial state, an array of time values at which the system state is calculated, and one or more operators, the time-dependent expectation value of each operator will be returned in the solver results. A minimal example is given in Listing~\ref{lst:spintime}.

\begin{lstlisting}[caption={Solving for the time-evolution of a spin-1/2 system subject to an external magnetic field oriented in the $+z$ direction and initially in the state $|+x\rangle$.},label={lst:spintime}]
t = arange(0,4*pi/omega,0.05)
expect_ops = [Sx,Sy,Sz]
psi0 = px
result = sesolve(Hz, psi0, t, expect_ops)
\end{lstlisting}

\begin{figure}
  \includegraphics[width=8.6cm]{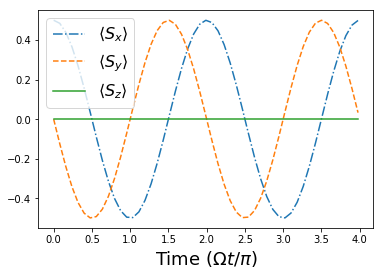}
  \caption{Spin-1/2 evolution.}
  \label{fig:spinplot}
\end{figure}

The result from running the \verb|sesolve()| is stored in the variable
\verb|result|, a specialized object defined within QuTiP. Among many
elements, it contains three arrays, each of which corresponds to one of the expectation
values passed in to the \verb|sesolve()| function. The first of these is
accessed as \verb|result.expect[0]|, which corresponds to $\langle
S_x\rangle(t)$ for each time value in the array \verb|t|. Summarizing the
results in a graph is demonstrated with the example code in
Listing~\ref{lst:spinplot} which generates Fig.~\ref{fig:spinplot}. We see $\langle
S_z \rangle=0$ as expected for a spin that will precess around the $+z$-axis.
Additionally, $\langle S_x \rangle$ and $\langle S_y \rangle$ oscillate out of phase
consistent with spin precession in the $xy$-plane.

\begin{lstlisting}[caption={Code used to generate Fig.~\ref{fig:spinplot}},label={lst:spinplot}]
labels = ['x','y','z']
style = {'x':'-.', 'y':'--', 'z':'-'}
for r,l in zip(result.expect,labels):
    plt.plot(t*omega/pi, r, style[l],
        label="$\langle S_%c \\rangle $" % l)

plt.xlabel("Time ($\Omega t/\pi$)", size=18)
plt.legend(fontsize=16)
\end{lstlisting}

\begin{figure}
  \includegraphics[width=8.6cm]{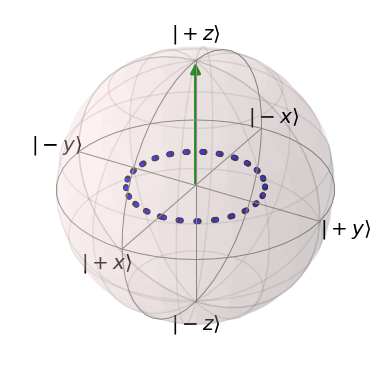}
  \caption{Bloch sphere visualization of spin-1/2 precession around a vertical B-field, indicated by the green vector. Blue dots are plotted at the coordinates ($\langle S_x \rangle(t)$,$\langle S_y \rangle(t)$,$\langle S_z \rangle(t)$)}
  \label{fig:bloch}
\end{figure}

To visualize this precession in a more intuitive way, we can take advantage of the Bloch sphere representation, implemented as \verb|Bloch()| in QuTiP. The following generates the Bloch sphere shown in Fig.~\ref{fig:bloch}.

\begin{lstlisting}[caption={Code to generate Fig.~\ref{fig:bloch}},label={lst:bloch}]
sx, sy, sz = result.expect
b = Bloch()
b.add_points([sx,sy,sz])
b.zlabel = ['$\\left|+z\\right>$',
            '$\\left|-z\\right>$']
b.ylabel = ['$\\left|+y\\right>$',
            '$\\left|-y\\right>$']
b.xlabel = ['$\\left|+x\\right>$',
            '$\\left|-x\\right>$']
b.add_vectors([0,0,1])
b.show()
\end{lstlisting}

Taking the sample code in Listing~\ref{lst:bloch} line-by-line, the three expectation values are unpacked from the \verb|result.expect| variable, a Bloch visualization object \verb|b| is created, we add the points described by the three components of the expectation value, set appropriate labels, and add an illustrative green arrow to indicate the direction of the magnetic field. Finally, \verb|b.show()| displays the Bloch visualization. In certain environments, a live 3D display of the Bloch sphere is available as \verb|Bloch3D()| with similar methods.\footnote{For more information, see \protect\url{qutip.org/docs/latest/guide/guide-bloch.html}}

\section{Harmonic Oscillator}
Moving to the QM simple harmonic oscillator (SHO) is a natural advancement in the undergraduate curriculum, and is also straightforward in QuTiP. While the SHO system is often taught from a wave mechanics perspective, it is more readily treated in QuTiP using the annihilation and creation operators, $\hat{a}$ and $\hat{a}^\dagger$, respectively. These operators are defined as \verb|destroy()| and \verb|create()|. Both take a single argument, \verb|N|, that is the dimension of the Hilbert space containing the operators. Formally, these are infinite-dimensional operators, which leads to our first encounter with the limitation of a numerical approach. A natural question, and one we encourage students to grapple with, is ``How big does N need to be?" The snarky answer is of course, ``big enough." But how do we evaluate that?

\begin{figure}
  \includegraphics[width=8.6cm]{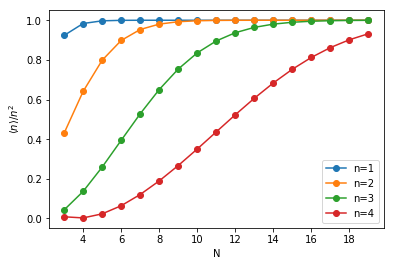}
  \caption{Plotting $\langle n \rangle/n^2$ vs. $N$ illustrates that small $N$ values truncates the Hilbert space and leads to numerical errors. These errors are reduced for larger values of $N$ but increasing $N$ will ultimately increase computation time. QuTiP is quite efficient with large vectors and matrices and we have not found this to be a limitation even when simulating large-dimension systems.}
  \label{fig:bigN}
\end{figure}

A short code sample in Listing~\ref{lst:bigN} illustrates one approach to answering this question more formally, the results of which are shown in Fig.~\ref{fig:bigN}. Here, we represent coherent states of $n$ quanta. For each state, we calculate the ratio $\langle n \rangle/n^2$. Quantum Mechanics rules that a coherent state will have $\langle n \rangle = n^2$ for all $n$ so this ratio should be 1 if the system is being represented accurately\cite{Beck:2012aa}. To test the accuracy of representation, several different coherent states (with differing $n$ values) are represented in $N$-dimensional Hilbert spaces. As $N$ increases, $\langle n \rangle/n^2$ goes to 1. For small $n$, the representation is accurate for $N=8$, but for $n=4$, errors persist even when the system is large $N=19$. In order to represent large-$n$ states, very large values of $N$ are required.

\begin{lstlisting}[caption={Code to generate Fig.~\ref{fig:bigN}},label={lst:bigN}]
# Plot the value of <n>/n^2 vs dimension N
for n in range(1,5):
    outn = []
    outN = []
    for N in range(3,20):
        a = destroy(N)
        psi = coherent(N,n)
        outN.append(N)
        outn.append( (psi.dag()*a.dag()*a*psi).data[0,0] )
    label_text = "n=%d" % n
    plt.plot(outN,real(outn)/n**2,"o-",label=label_text)

plt.xlabel("N")
plt.ylabel("$\\langle n \\rangle / n^2$")
plt.title("How big does N need to be?");
plt.legend()
\end{lstlisting}

In addition to functions described above, there are several built-in functions that are useful in the context of the harmonic oscillator. Three standard states (coherent, thermal, and Fock) are available as both vector (ket) representations and density matrix representations as shown in Listing~\ref{lst:sho_states}. In each of these, the $N$ parameter specifies the size of the matrix used to represent the state (as above), and the $n$ parameter indicates the occupation number for the Fock state, or the expectation value $\langle n \rangle$ in the thermal state. Coherent states have \verb|alpha|$=\alpha$ as the second parameter where $\alpha$ is the eigenvalue of the coherent state: $\hat{a}|\alpha\rangle = \alpha|\alpha\rangle$.

\begin{lstlisting}[caption={Built-in states},label={lst:sho_states}]
coherent(N,alpha)
fock(N,n)
coherent_dm(N,alpha)
fock_dm(N,n)
thermal_dm(N,n)
\end{lstlisting}

\subsection{Visualizing SHO states}
\begin{figure}
  \includegraphics[width=8.6cm]{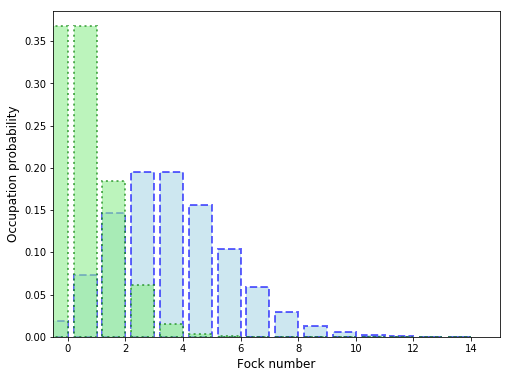}
  \caption{The Fock distribution for a coherent state with $\alpha=2$ (blue,dashed) and with $\alpha=1$ (green,dotted).}
  \label{fig:fock_dist}
\end{figure}

There are a variety of visualization methods for SHO states in QuTiP because the package has been designed for use in quantum optics where a single mode of the electromagnetic field may be quantized as a harmonic oscillator. The Fock distribution indicates the probability of measuring a given number of quanta (i.e. photons) for a particular state. The coherent state is not an energy eigenstate and thus contains contributions from a range of Fock states as shown in Fig.~\ref{fig:fock_dist}. This distribution illustrates the key differences between Fock states and coherent states, and lends visual understanding to the sum in the definition of the coherent state: $$|\alpha\rangle = e^{-|\alpha|^2/2}\sum_{n=0}^{\infty}\frac{\alpha^n}{\sqrt{n!}}|n\rangle.$$

\begin{figure}
  \includegraphics[width=8.6cm]{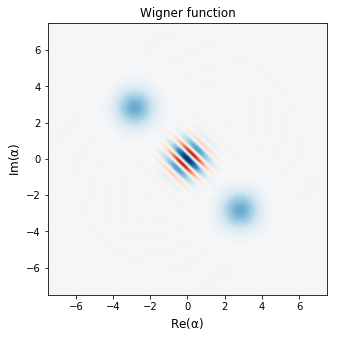}
  \caption{The Wigner function for a sum of two coherent states with $\alpha=\pm2\mp2i$. Blue values are positive, red are negative.}
  \label{fig:wigner}
\end{figure}

The Wigner function is a pseudo-probability distribution often used to visualize the states of a harmonic oscillator with a particular application to the modes of light in quantum optics.\cite{Leonhardt:1997aa} The Wigner function is analogous to a phase-space distribution for classical systems. The quantum version stretches this interpretation with the possibility of containing negative values.\footnote{Negative values in the Wigner function justify the description as a pseudo-probability distribution.} Quantum interference effects are the source of such values as we show here. The QuTiP method \verb|wigner(psi,x,y)| computes the Wigner function for state \verb|psi| over coordinate arrays \verb|x| and \verb|y|, or the \verb|plot_wigner(psi)| function handles plotting and coordinate generation automatically. Listing~\ref{lst:wigner} gives a simple example that generates Fig.~\ref{fig:wigner}. The state visualized here is a sum of two coherent states with different values of $\alpha=\pm2\mp2i$. Note that python defines the complex constant $i$ as \verb|1j| and a general complex value such as $2i$ as \verb|2j|. The blue values in Fig.~\ref{fig:wigner} are positive and the red values are negative. The  fringes in the region between the two gaussian portions are a signature of quantum interference that has no classical analogy in phase space. These superpositions of coherent states are known as Schr\"odinger cat states and have a variety of applications.\cite{Gerry:1997aa}

\begin{lstlisting}[caption={Plotting the Wigner function},label={lst:wigner}]
plot_wigner(coherent(25,-2+2j)+coherent(25,2-2j))
\end{lstlisting}

\section{Advanced Topics}
\label{sec:advanced}
QuTiP gives students the ability to extend the fundamental explorations presented above and tackle problems that are closer to the cutting edge. All of the methods, techniques, and quantum objects that they have learned to use in simpler problems maintain their relevance.

\begin{figure}[ht!]
\centering
\includegraphics{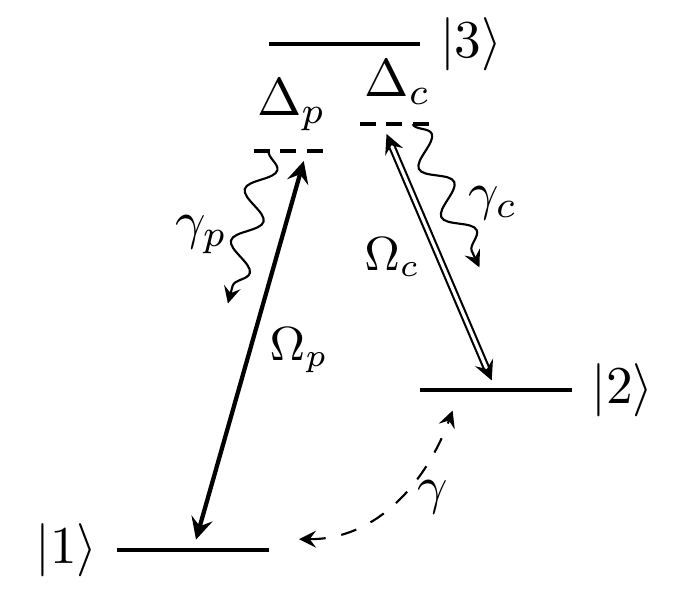}
\caption{Atomic energy levels in EIT. The three atomic states are designated $|1\rangle$, $|2\rangle$, and $|3\rangle$. In EIT, $|2\rangle$ is a meta-stable state such that the coherence between $|1\rangle$ and $|2\rangle$ is long-lived. Excited state $|3\rangle$ undergoes spontaneous emission decay to $|1\rangle$ and $|2\rangle$ with rates $\gamma_p$ and $\gamma_c$ respectively.}
\label{fig:EITlevels}
\end{figure}

One example of a more advanced application of QuTiP is the study of Electromagnetically-Induced Transparency (EIT). A minimal example of an EIT system is presented here for completeness, further details can be found in other recent literature. We follow the theoretical treatment used by Erickson \cite{Erickson:2012aa} and Fleischhauer \cite{Fleischhauer:2005aa}. The most relevant features of this system are captured in the interaction Hamiltonian which describes the coupling between two classical beams of light (the probe and coupling fields) and a quantum-mechanical three-level atom. The Hamiltonian, after several standard approximations, is given as:\footnote{The tilde indicates this operator has undergone unitary transformation into a corotating frame as described in Ref.~\onlinecite{Erickson:2012aa}}

\begin{equation}
    \tilde{H}_{EIT} = -\frac{\hbar}{2}
  \begin{bmatrix}
    0 & 0 & \Omega_p \\
    0 & 2(\Delta_p - \Delta_c) & \Omega_c\\
    \Omega_p & \Omega_c & 2\Delta_p
  \end{bmatrix}
  \label{eqn:Heit}
\end{equation}

We point out that Refs.~\onlinecite{Erickson:2012aa} and \onlinecite{Fleischhauer:2005aa} differ in their sign convention for the field detunings $\Delta_p$ and $\Delta_c$. We follow the definition in Ref.~\onlinecite{Erickson:2012aa} where $\Delta_p=\omega_p - (\omega_3 - \omega_1)$ and $\Delta_c=\omega_c - (\omega_3 - \omega_2)$ with bare-atom energies $E_n = \hbar\omega_n$.

For the example here, we set several parameters for the system. Of the parameters included in the Hamiltonian, we set the coupling field detuning $\Delta_c=0$, probe-field Rabi frequency $\Omega_p=0.01$, and the coupling-field Rabi frequency $\Omega_c=1.0$. Additionally, we define units such that the relevant decay rates from the excited state: $\gamma_p = \gamma_c = 1$ and the ground state coherence decay rate $\gamma=0.05$. These parameters are defined in python as shown in Listing~\ref{lst:eitparameters}.

\begin{lstlisting}[caption={EIT parameters},label={lst:eitparameters}]
gamma_p = 1  # decay rate on probe transition
gamma_c = 1  # decay rate on coupling transition
gamma = 0.05  # ground coherence decay
deltaC = 0  # coupling field detuning
omegaP = 0.01  # Rabi freq. for probe
omegaC = 1.0  # Rabi freq. for control
\end{lstlisting}

To realize a numerical version of this system, we make use of the QuTiP solver \verb|steadystate| which computes the steady state given the Hamiltonian and a set of collapse operators. We use the Hamiltonian given in Eqn.~\ref{eqn:Heit} and create a set of collapse operators based on the decay rates for each atomic state. The first step in defining this system is to create basis states for each of the three atomic energy levels. We name these states according to Fig.~\ref{fig:EITlevels}. Additionally, we define relevant projection operators $\hat{\sigma}_{ij}=|i\rangle\langle j|$ $(i,j = 1,2,3)$.

\begin{lstlisting}[caption={EIT levels and atomic projection operators},label={lst:eitproj}]
# Define the levels and atomic projection operators
one, two, three = three_level_basis()
sig_11 = one * one.dag()
sig_22 = two * two.dag()
sig_33 = three * three.dag()
sig_13 = one * three.dag()
sig_23 = two * three.dag()
sig_12 = one * two.dag()
\end{lstlisting}

The collapse operators have the form $C_n=\sqrt{\gamma_n}A_n$ where $\gamma_n$ is the decay rate and $A_n$ is the operator that describes the decay.\cite{Johansson:2012aa} For an EIT system, the three collapse operators correspond to coherence decay for the three coherences in the density matrix. The probe transition ($\hat{\sigma}_{13}$), coupling transition ($\hat{\sigma}_{23}$), and two ground states ($\hat{\sigma}_{12}$) decay with rates given above in python as \verb|gamma_p|, \verb|gamma_c|, and \verb|gamma|, respectively. We define the associated collapse operators and create a python list that contains them as shown in Listing~\ref{lst:eitcollapse}.

\begin{lstlisting}[caption={EIT collapse operators},label={lst:eitcollapse}]
c1 = np.sqrt(gamma_p)*sig_13  # 1-3 coherence decay
c2 = np.sqrt(gamma_c)*sig_23  # 2-3 coherence decay
c3 = np.sqrt(gamma)*sig_12  # 1-2 coherence decay
collapse = [c1,c2,c3]
\end{lstlisting}

The remaining parameter is the probe-field detuning ($\Delta_p$). For this, we define a list of values in order to solve for the steady state of the system for each value of $\Delta_p$. Using the \verb|linspace| function again, we create a list of values \verb|deltalist| as well as an empty list that will hold the results as shown in Listing~\ref{lst:detuning}

\begin{lstlisting}[caption={EIT probe detunings},label={lst:detuning}]
# create list of delta_p values
deltalist = np.linspace(-3,3,200)

# empty list to save results
chi = []  # susceptibility
\end{lstlisting}

While the full details are beyond the scope of this paper, the electric susceptibility associated with the probe field is directly proportional to the coherence $\rho_{13}$.\cite{Boyd:2002aa} Calculating the steadystate value of $\rho_{13}$ will therefore provide insight into the optical properties of the system as $\Delta_p$ is varied near resonance. Listing~\ref{lst:eitloop} uses a \verb|for| loop to calculate the steadystate density matrix \verb|rho_ss| for each value of \verb|deltalist|. Additionally, we compute the expectation value of the projection operator $\hat{\sigma}_{13}$ which gives a measure of the electric susceptibility $\chi$. These values are appended to a list \verb|chi| at each iteration of the loop making them available for plotting in Listing~\ref{lst:eitgraphs}.

\begin{figure}
  \includegraphics[width=8.6cm]{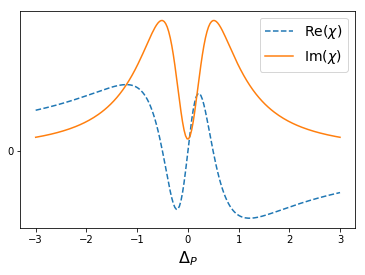}
  \caption{The real and imaginary parts of the electric susceptibility correspond to the refractive index ($n-1$) and the absorption respectively. The steep positive slope of $n$ vs. $\Delta_p$ indicates a frequency range for which the system exhibits strong dispersion associated with low group velocity, a phenomena known as slow light.\cite{Boyd:2002aa}}
  \label{fig:eit}
\end{figure}

\begin{lstlisting}[caption={EIT calculation loop},label={lst:eitloop}]
for deltaP in deltalist:
    H = -1/2*Qobj([[0,0,omegaP],[0,2*(deltaP - deltaC),omegaC],[omegaP, omegaC, 2*deltaP]])
    rho_ss = steadystate(H,collapse)
    chi.append(expect(sig_13,rho_ss))
\end{lstlisting}

Finally, we can plot the real and imaginary parts of the resulting $\chi$ values to demonstrate two key features of an EIT system: transparency near resonance, and steep normal dispersion. The solid (orange) curve in Fig.~\ref{fig:eit} shows the absorption which is nearly zero on resonance ($\Delta_p=0$). This absorption dip indicates a frequency range that experiences transparency. To confirm this, one can re-run the simulation with a weak coupling field ($\Omega_c \leq 0.1$). For such conditions, a single resonance absorption peak results, and at $\Delta_p=0$ the absorption is high instead of low (as expected for a resonant field).

\begin{lstlisting}[caption={Graphing EIT results, shown in Fig.~\ref{fig:eit}},label={lst:eitgraphs}]
plt.plot(deltalist,np.real(chi),label="Index n")
plt.plot(deltalist,np.imag(chi),label="Absorption")
plt.legend()
plt.xlabel("$\Delta_p$")
plt.yticks([])
\end{lstlisting}

The second notable feature of EIT is a region of strong normal dispersion (refractive index increases with frequency). Such dispersion is known to cause low group velocity for pulses of resonant light.\cite{Boyd:2002aa} Normal dispersion is evident near resonance in the dashed (blue) curve of Fig.~\ref{fig:eit}.

\section{Suggested Problems}

As a full-featured software package, QuTiP offers many opportunities for extended work. Many in-depth examples already exist as part of the documentation, but we suggest a few starting points here that are more consistent with advanced undergraduate work.

\subsection{Computing and gates}

Quantum computing is moving from the realm of fundamental physics toward engineering applications. Companies like IBM and Intel have teams devoted to building quantum computers. While currently far from market-ready, there is rapid development in the number of qubits per system.

In this brief introduction, we will demonstrate the features of quantum gates that connect single-spin evolution to qubit gates and basic computational methods in the context of tools available in the QuTiP package.

Quantum circuits consist of a set of qubits, represented by horizontal lines, not unlike a music staff. Each qubit evolves in time by following it's line to the right. Along this trajectory, gates may act on one or more qubits resulting in changes to the state of each individual qubit, and therefore changes to the state of the system as a whole. There are similarities to electronic circuits here, but there are also several key differences. One is that this system, as a fundamentally quantum system, is not deterministic (unlike a basic battery, wire, bulb circuit). This is the primary advantage of quantum computing: many simultaneous parallel computations using a finite set of states.

In this section, we present basic examples of quantum circuits, using the QuTiP tools, with suggestions for taking the topic further. The first step is to create an object that represents the quantum circuit. This object's methods will allow us to assemble a complete circuit and perform calculations for the entire system. The \verb|add_gate| method takes several arguments: \verb|gate|, \verb|targets|, \verb|controls|, \verb|arg_value|, \verb|and arg_label|. Not all gates require controls, so these can be left as None or skipped by using named arguments as shown in the second \verb|add_gate| in Listing~\ref{lst:qcXY}.

We will start with two simple gates, in fact we have seen these in a different context earlier. The RY gate rotates a qubit around the Y axis, and the RX gate rotates a qubit around the X axis. The amount of rotation is specified in the \verb|add_gate|| method, as are the target qubit, and a label for the gate (which will appear in the figure generated by QuTiP).

\begin{lstlisting}[caption={A 2-qubit circuit with RX and RY gates.},label={lst:qcXY}]
N=2
qc = QubitCircuit(N)
qc.add_gate("RY", 0, arg_value=pi/2, arg_label=r"\pi/2")
qc.add_gate("RX", 1, arg_value=pi/2, arg_label=r"\pi/2")
qc.png
\end{lstlisting}

Finally, we compute the full propagator (the matrix product that represents the full sequence of gates). This is done in two steps, first, generate a list of the propagators \verb|U_list| and then calling the function \verb|gate_sequence_product| to compute the product:

\begin{lstlisting}[caption={}]
U_list = qc.propagators()
U1 = gate_sequence_product(U_list)
\end{lstlisting}

To apply this circuit to a specific system, we prepare an initial state, in this case two ground-state qubits, and act on the initial state with the gate sequence product:

\begin{lstlisting}[caption={}]
initial = tensor(basis(N,0),basis(N,0))
final = U1*initial
\end{lstlisting}

We can explore the effect of these gates on each individual qubit. To do this, we use the Bloch representation again, and generate a Bloch vector for the initial and final spins (before and after application of the gate).

\begin{lstlisting}[caption={Bloch sphere representation of Ry gate}]
before = Bloch()
after = Bloch()
before.add_states(initial.ptrace(0)) # look at only qubit 0 with the ptrace
after.add_states(final.ptrace(0))
before.show()
after.show()
\end{lstlisting}

\begin{figure}
  \includegraphics[width=4cm]{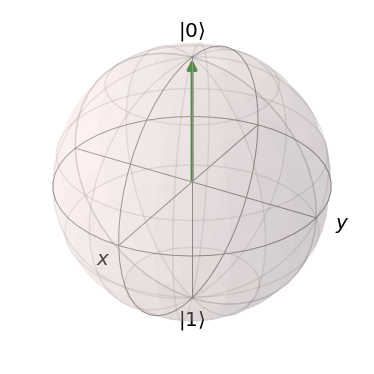}
  \includegraphics[width=4cm]{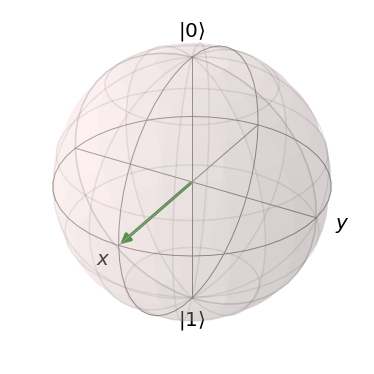}
  \caption{Bloch sphere representation of qubits before (left) and after (right) application of the $R_y(\pi/2)$ gate.}
  \label{fig:qcXY}
\end{figure}

\begin{lstlisting}[caption={Bloch sphere representation of Rx gate }]
before = Bloch()
after = Bloch()
before.add_states(initial.ptrace(1)) # look at only qubit 1 with the ptrace
after.add_states(final.ptrace(1))
before.show()
after.show()
\end{lstlisting}

\begin{figure}
  \includegraphics[width=4cm]{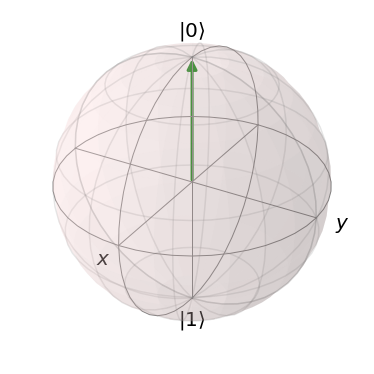}
  \includegraphics[width=4cm]{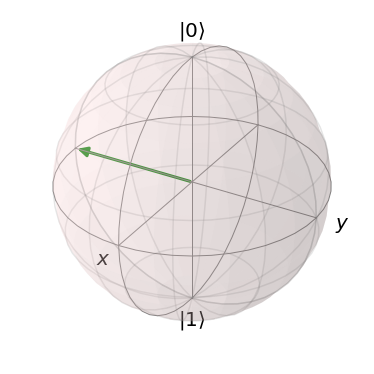}
  \caption{Bloch sphere representation of qubits before (left) and after (right) application of the $R_x(\pi/2)$ gate.}
  \label{fig:qcXY}
\end{figure}

\subsection{Entanglement: Bell state generation and tests}
A common requirement for quantum circuits is to have entangled qubits. To achieve this from a known initial state we make use of the Hadamard gate SNOT and the controlled-not gate CNOT:

\begin{lstlisting}[caption={}]
bellgen=QubitCircuit(2)
bellgen.add_gate("SNOT", 1)
bellgen.add_gate("CNOT", 0, 1)
bg_gates = gate_sequence_product(bellgen.propagators())
initial = tensor(basis(2,0),basis(2,0))
final = bg_gates*initial
\end{lstlisting}

We can generate the density matrix for the initial state via \verb|initial*initial.dag()| and for the final state via \verb|final*final.dag()| which produces the expected results (respectively):
\[
\begin{pmatrix}
1 & 0 & 0 & 0 \\
0 & 0 & 0 & 0 \\
0 & 0 & 0 & 0 \\
0 & 0 & 0 & 0
\end{pmatrix}
\]
\[
\begin{pmatrix}
0.5 & 0 & 0 & 0.5 \\
0 & 0 & 0 & 0 \\
0 & 0 & 0 & 0 \\
0.5 & 0 & 0 & 0.5
\end{pmatrix}
\]

As a final check, we can evaluate the concurrence of the initial and final states as a measure of entanglement\cite{Wootters:1998aa}.

\begin{lstlisting}[caption={}]
  print("before: C=",concurrence(initial))
  print("after: C=",concurrence(final))
\end{lstlisting}

Which reports $C=0$ for the initial state, and $C=0.9999$ for the final state, thus demonstrating that the combination of SNOT and CNOT gates serve to create two entangled qubits from one known initial state.

\subsection{Atomic Physics}
In addition to serving as the primary package used for a computation, QuTiP can be built into more advanced and involved calculations. As an example of this, we will illustrate the use of the \verb|maxwellbloch| package \cite{Ogden:2016aa}. There are many excellent examples available by the package author, and we merely include one here to serve as an introduction. Given that this package is quite flexible, we must first explain how the simulations are defined. A JavaScript Object Notation (JSON) string is used to define the relevant system properties.\footnote{While not written in Java, this file format has become a common standard as it is both easy to read (by humans) and simple to parse in software.} For simplicity, we consider the same EIT system described in Section~\ref{sec:advanced}. The JSON file for this system is shown in Listing~\ref{lst:eit_json}.

\begin{lstlisting}[caption={JSON file for EIT with no coupling\footnote{See Supplemental Materials for more details on using this example.}},label={lst:eit_json}]
mb_solve_json = """
{
  "atom": {
    "num_states": 3,
    "decays": [
      { "channels": [[0,1], [1,2]],
        "rate": 0.0
      }
    ],
    "fields": [
      {
        "coupled_levels": [[0, 1]],
        "detuning": 0.0,
        "detuning_positive": true,
        "label": "probe",
        "rabi_freq": 1.0e-3,
        "rabi_freq_t_args":
          {
             "ampl": 1.0,
             "centre": 0.0,
             "fwhm": 1.0
          },
        "rabi_freq_t_func": "gaussian"
      },
      {
        "coupled_levels": [[1, 2]],
        "detuning": 0.0,
        "detuning_positive": false,
        "label": "coupling",
        "rabi_freq": 0.0,
        "rabi_freq_t_args":
        {
          "ampl": 1.0,
          "fwhm": 0.2,
          "on": -1.0,
          "off": 9.0
          },
        "rabi_freq_t_func": "ramp_onoff"
      }
    ]
  },
  "t_min": -2.0,
  "t_max": 10.0,
  "t_steps": 120,
  "z_min": -0.2,
  "z_max": 1.2,
  "z_steps": 140,
  "z_steps_inner": 10,
  "num_density_z_func": "square",
  "num_density_z_args": {
    "on": 0.0,
    "off": 1.0,
    "ampl": 1.0
  },
  "interaction_strengths": [100.0, 100.0],
  "method": "mesolve"
}
"""
\end{lstlisting}

The JSON file contains a hierarchy of parameters that describe the system. The top level describes the atom. It's parameters are \verb|num_states|, \verb|decays|, and \verb|fields|. The three states of this atom are designated 0,1,2. There are two decay channels, both assigned a rate of zero for the time being. The first is between levels 0 and 1, the second between levels 1 and 2. We define two fields, probe and coupling, as before in the EIT example. Each field is assigned to a pair of coupled levels and given a number of parameters. We set the detuning to zero, assign a label, and specify the Rabi frequency. This final parameter has many options in order to describe time-varying fields. For the probe field, we set the amplitude to 1e-3 and set the Rabi frequency to follow a Gaussian function in time (with amplitude 1.0 and FWHM 1.0). Note: the units on these parameters are scaled for simplicity.

The coupling field is set in a similar way, although the Rabi frequency is either 0.0 (for no coupling) or 5.0 for coupling. The time-function for the coupling fields is specified as \verb|ramp_onoff| with amplitude 1.0, FWHM of 2.0, on-time of -1.0 and off-time of 9.0 (which is well after the pulse has left the atomic medium.

\begin{figure}
  \includegraphics[width=8.6cm]{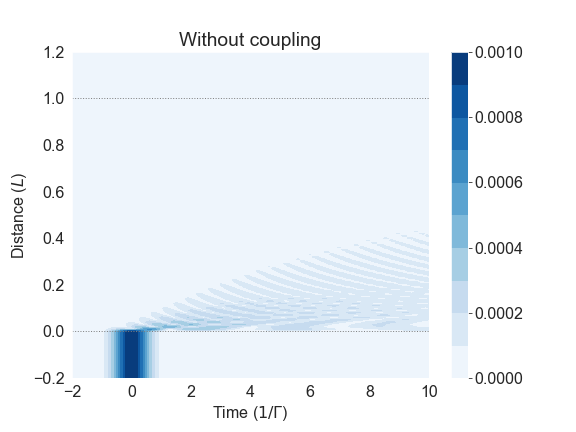}
  \caption{Three-level lambda system with no coupling field. Pulse absorption is complete.}
  \label{fig:mb-no-coupling}
\end{figure}

\begin{figure}
  \includegraphics[width=8.6cm]{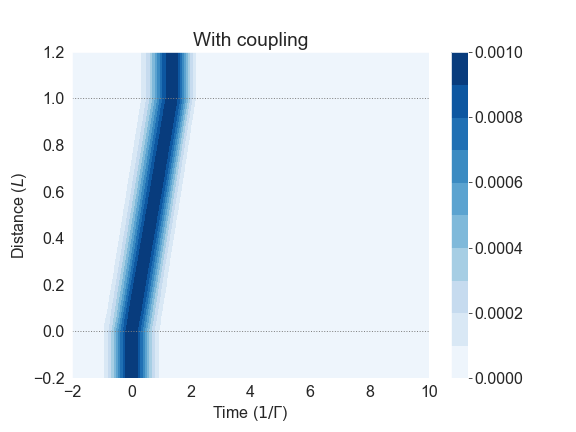}
  \caption{Three-level lambda system with coupling field. Pulse is delayed and completely transmitted.}
  \label{fig:mb-w-coupling}
\end{figure}

We see complete absorption of the pulse in the case with no coupling. This is identical to the situation with a pulse that is on-resonance in a two-level system. When the coupling field is enabled, the pulse propagates, albeit slowly, through the material. This is a demonstration of two prominent features of EIT: transparency and slow-light. Transparency is evident from the pulse maintaining amplitude, and slow-light is shown by the slight diagonal path of the pulse in the moving reference frame. A pulse traveling at the speed of light would remain centered at 0 on the horizontal axis, but slow-light lags behind and exits the medium at a later time (in this case roughly 1/$\Gamma$).

\section{Conclusion}

The QuTiP python package provides a robust numerical framework for the study of quantum systems using conventions and notation that make adding computational exercises to an undergraduate course straightforward. In addition to the systems discussed here, the QuTiP documentation and online examples demonstrate the broad capabilities of this package and provide interested users with a variety of starting points in a wide range of topical areas. The newest features of QuTiP provide for the study of quantum control algorithms that are central to the emerging field of quantum computing.

\begin{acknowledgments}

The author thanks T. Ogden, J.D. Lett, E. LeVally, and M. Brown for providing valuable feedback on early drafts of this manuscript, and gratefully acknowledges support from NSF award 1506049.

\end{acknowledgments}

\bibliography{AJPqutip}

\end{document}